\documentclass{sig-alternate-05-2015}
\usepackage{times} 

\usepackage{amsmath}
\usepackage{amsfonts}
\usepackage{amssymb}
\usepackage{graphicx}
\usepackage{morefloats}
\usepackage{caption}
\usepackage{subfigure}
\usepackage{epstopdf}
\usepackage{color}
\usepackage{pbox}
\usepackage{array}
\usepackage{multirow}
\usepackage{graphicx}
\usepackage[sort,nocompress]{cite}
\usepackage{url}

\usepackage{breakurl} 
\usepackage[breaklinks]{hyperref}
\usepackage[boxed,vlined]{algorithm2e}
\usepackage{paralist} 
\usepackage{listings} 

\newcommand{\bigcell}[2]{\begin{tabular}{@{}#1@{}}#2\end{tabular}}

\usepackage{tikz}
\begin{document}

\title{Me, Myself and My Killfie:\\Characterizing and Preventing Selfie Deaths}

\author{
Hemank Lamba$^1$, Varun Bharadhwaj$^3$, Mayank Vachher$^2$,\\
Divyansh Agarwal$^2$, Megha Arora$^1$, Ponnurangam Kumaraguru$^2$	\\
\\
$^1$Carnegie Mellon University, USA 	\\
\{hlamba@cs,marora@andrew\}.cmu.edu	\\
$^2$Indraprastha Institude of Information Technology, Delhi, India	\\
\{mayank13059,divyansha,pk\}@iiitd.ac.in\\
$^3$National Institute of Technology, Tiruchirappalli	\\
var6595@gmail.com
}

\maketitle


\begin{abstract}
\noindent 

Over the past couple of years, clicking and posting selfies has become a popular trend. However, since March 2014, 127 people have died and many have been injured while trying to click a selfie. Researchers have studied selfies for understanding the psychology of the authors, and understanding its effect on social media platforms. In this work, we perform a comprehensive analysis of the selfie-related casualties and infer various reasons behind these deaths. We use inferences from incidents and from our understanding of the features, we create a system to make people more aware of the dangerous situations in which these selfies are taken. We use a combination of text-based, image-based and location-based features to classify a particular selfie as dangerous or not. Our method ran on 3,155 annotated selfies collected on Twitter gave 73\% accuracy. Individually the image-based features were the most informative for the prediction task. The combination of image-based and location-based features resulted in the best accuracy. We have made our code and dataset available at \url{http://labs.precog.iiitd.edu.in/killfie}.

\end{abstract}

\section{Introduction}
\label{sec:intro}








With the rise in the amount and type of content being posted on social media, various trends have emerged. In the past, social media trends like memes~\cite{gharan2010memes, leskovec2009memetracker, simmons2011memes}, social media advertising~\cite{miller2010social}, firestorm~\cite{lamba2015tempest}, crisis event reporting~\cite{sakaki2010earthquake, sakaki2011tweet}, and much more have been extensively analyzed. Another trend that has emerged over social media in the past few years is of clicking and uploading selfies. According to Oxford dictionary, a selfie is defined as \textit{a photograph that one has taken of oneself, typically one taken with a smart phone or web cam and shared via social media}~\cite{selfie_def}. A selfie can not only be seen as a photographic object that initiates the transmission of the human feeling in the form of a relationship between the photographer and the camera, but also as a gesture that can be sent via social media to a broader population~\cite{senft2015selfies}. Google estimated that a staggering 24 billion selfies were uploaded to Google Photos in 2015~\cite{Kennemer14selfie}. The selfie trend is popular with millennials (ages 18 to 33). Pew research center found that around $55$\% of millennials have posted a "selfie" on a social media service~\cite{pew2014selfie_report}. The popularity of selfie trend is so massive that "selfie" was declared as the word of the year in 2013 by Oxford Dictionary~\cite{oxford2013word}. The virality of the selfie culture has also been known to cause service interruptions on popular social media platforms. For instance, the selfie taken by Ellen Degeneres, a popular television host, at the Academy Awards brought down Twitter website due to its immense popularity~\cite{Ellen2014Tweet}.

Selfies have proved instrumental in revolutionary movements~\cite{brager2015selfie}, and have also known to help election candidates increase their popularity~\cite{baishya2015namo}. Many researchers have studied selfies for understanding psychological attributes of the selfie authors~\cite{lakshmi2015selfie, qiu2015does}, investigating the effect of selfies on social protests~\cite{brager2015selfie}, understanding the effect of posting selfies on its authors~\cite{senft2015selfies}, dangerous incidents and deaths related to selfies~\cite{bhogesha2016death, howes2015let, subrahmanyam2016selfie} and using computer vision methods to interpret whether a given image is a selfie or not~\cite{carmeanselfie}.

\begin{figure}
\centering
\begin{tabular}{lr}
\scalebox{0.18}{\includegraphics[angle=0,
origin=c]{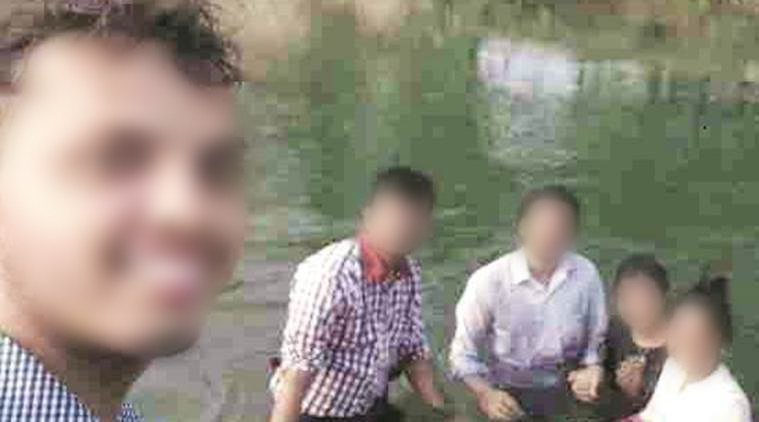}} &
\scalebox{0.23}{\includegraphics[angle=0,
origin=c]{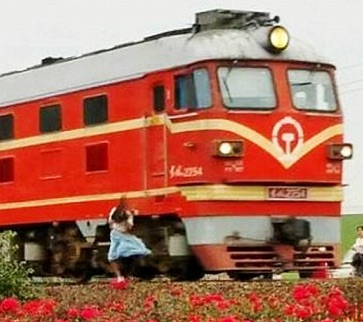}}
\end{tabular}
\caption[Dangerous Selfies]{Left: Selfie took by a group of individuals shortly before they drowned in the lake.
Right: Photograph of a girl taking a selfie on train tracks immediately before a train hit her.}
\label{fig:selfie_true_images}
\end{figure}

Clicking selfies has become a symbol of self-expression and often people portray their adventurous side by uploading crazy selfies~\cite{dangerous_selfies}. This has proved to be dangerous~\cite{subrahmanyam2016selfie, howes2015let, bhogesha2016death}. Keeping in mind the hazardous implications of taking selfies at dangerous locations, Russian authorities came up with public posters, indicating the dangers of taking selfies~\cite{russiaposters2015}. Similarly, Mumbai police recently classified $16$ zones across Mumbai as no-selfie zones~\cite{mumbainoselfie2016}. Through the process of data collection, we found $127$ people have been killed since 2014 till September 2016 while attempting to take selfies. From $15$ casualties in 2014 and $39$ in 2015, the death toll due to selfies has reached $73$ till September 2016. It has been reported that the number of selfie deaths in 2015 was more than the number of deaths due to shark attacks~\cite{sharks15}. Some of the selfies that led to casualties are shown in the Figure~\ref{fig:selfie_true_images}. Given the influence of selfies and the significant rise in the number of deaths and injuries reported when users are taking selfies, it is important to study these incidents in detail and move towards developing a technology which can help reduce the number of selfie casualties.

\begin{figure*}[t]
\centering
\includegraphics[scale=0.3]{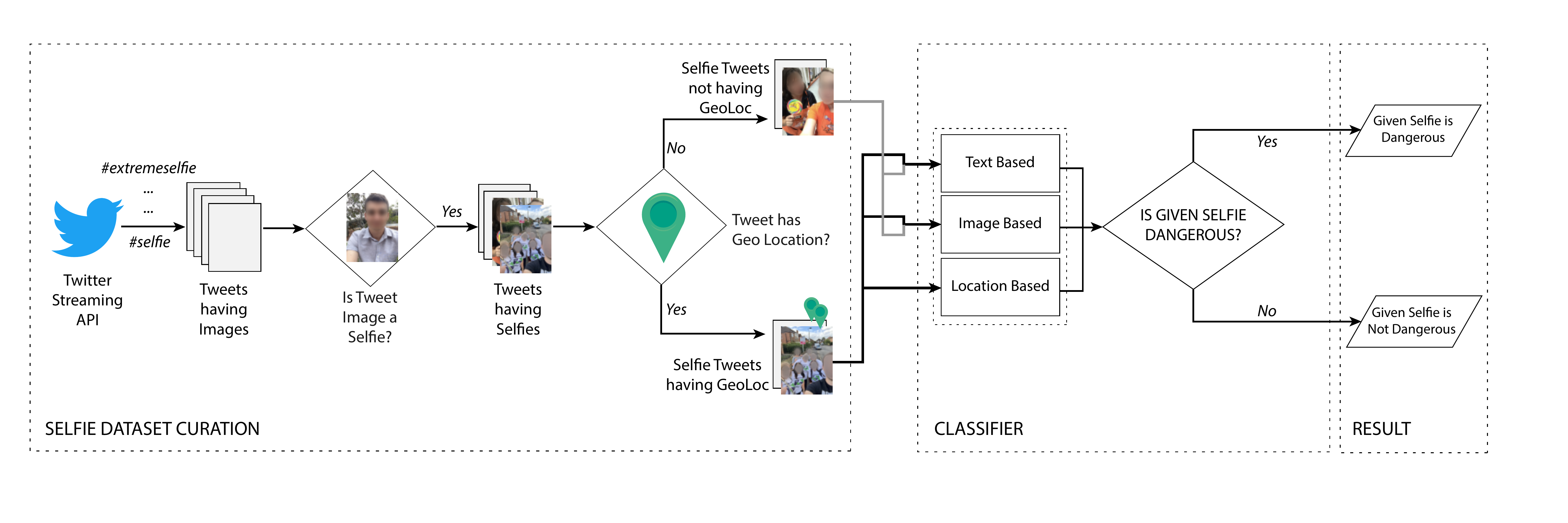}
\caption{A brief overview of our approach - Tweets tagged with a geolocation are analyzed using text, location and image-based features. Whereas tweets without a geolocation are analyzed only using text and image-based features.}
\label{fig:Sys_Arch}
\end{figure*}

 %
 %
In this paper, we characterize the demographics and analyze reasons behind selfie deaths; based on the obtained insights, we propose features which can differentiate potentially dangerous selfie images from the non-dangerous ones. Our methodology is briefly explained in Figure~\ref{fig:Sys_Arch}. Specifically, the major contributions of the paper are as follows:

\begin{itemize}
\item{\textbf{Data Characterization:} We do a thorough analysis of the selfie casualties, and provide insights about all the previous fatal selfie-related incidents.}
\item{\textbf{Feature Identification}: We propose features that are easily extractable from the social media data and learn signals which determine if a particular selfie is dangerous.}
\item{\textbf{Discriminative Model}: We present a model that based on the proposed features can differentiate between dangerous selfies and non-dangerous selfies.}
\item{\textbf{Real World Data}: We test our given approach on a real-world dataset collected from a popular social media website. We also test the efficacy of our approach in absence of certain features, a situation which is possible while working on such real datasets.}
\end{itemize}
Furthermore, we believe our contributions could lead to generation of tools or treatments that can have a significant impact on reducing the number of selfie deaths.\\
\textbf{Reproducibility:} More detailed analysis of the selfie deaths is shown on our web page\footnote{\url{http://labs.precog.iiitd.edu.in/killfie/}}, and our code and the dataset is also available for download.

\section{Related Work}
\label{sec:background}
%
%
%
%
%
%

The trend and culture of posting selfies on social media have been investigated widely over the past few years. The popularity of selfies being posted on online social media has drawn a lot of researchers from different fields to study the various aspects of the selfie trend. We present the relevant work from major fields in this section.

\textbf{The impact of selfies}: Brager et al. studied the effect of a particular selfie on playing a part in a revolutionary movement~\cite{brager2015selfie}. The authors specifically analyzed death of a young teenager in Lebanon who died moments after taking a selfie near a golden SUV, that blew up. His death and the specific selfie stirred the Western news media and spectators, revolutionizing the movement - \#NotAMartyr over the Internet. The authors argued that the practice of selfie-taking made the young boy's story legible as a subject of grievance for the Western social media audience. Porch et al. analyzed how the selfie trend has affected women's self-esteem, body esteem, physical appearance comparison score, and perception of self~\cite{porch2015society}. Baishya et al. found the effect of selfies by candidate prime minister in Indian general elections was significant towards his victory~\cite{baishya2015namo}. Lim et al. suggested that insights into the selfie phenomenon can be understood from socio-historical, technological, social media, marketing, and ethical perspectives~\cite{lim2016understanding}.

\textbf{Psychology Studies}: Qiu et al. analyzed the correlations between selfies and the personalities according to Big Five personality test of the participants~\cite{qiu2015does}. Authors used signals such as camera height, lips position and the portrayed emotion to make predictions about their emotional positivity, openness, neuroticism and conscientiousness. Li et al. proposed that people taking selfies have narcissistic tendencies and the selfie-takers use selfies as a form of self-identification and expression. The role of selfies was also analyzed in making the selfie-taker a journalist who posts images on social media after witnessing events~\cite{koliska2015selfies}. Senft et al. analyzed the role that selfies play in affecting the online users. It further shows how selfie as a medium has a narcissistic or negative effect on people~\cite{senft2015selfies}.

\textbf{Dangers of Selfie}: An important theme, which is directly related to our paper is work related to the dangers that trend of selfie taking puts a selfie-taker in. Lakshmi et al. explain how the number of likes, comments and shares they get for their selfies are the social currency for the youth. The desire of getting more of this social currency prompts youth to extreme lengths~\cite{lakshmi2015selfie}. Flaherty et al.~\cite{flaherty2016selfie} and Bhogesha et al.~\cite{bhogesha2016death} talk about how selfies have been a risk during international travel. Howes et al. analyzed the selfie trends as a cultural practice in the contemporary world~\cite{howes2015let}. Authors particularly analyzed the case of spectators clicking selfies in the sport of cycling. The spectators wanted to capture the moment but ended up in obstructing the path of cyclists, leading to crashes. Subrahmanyam et al. work is the closest to ours discussing the dangers of taking a selfie~\cite{subrahmanyam2016selfie}. Authors also provided statistical data about the number of deaths and injuries. A noble initiative \#selfietodiefor\footnote{\url{http://www.selfietodiefor.org/}} has been posting about the dangers of taking a selfie in a risky situation. They use Twitter handle @selfietodiefor for sending out awareness tweets and news stories related to selfie deaths. 

Besides all the above-mentioned areas, researchers have also tried to distinguish selfies from other images by use of automated methods~\cite{carmeanselfie}. A project called \textit{Selfie City} has been investigating the style of selfies in five cities across the world~\cite{women_men_selfies}. Using the dataset collected, they explored the age distribution, gender distribution, pose distribution and moods in all of the selfies collected. Researchers have also explored the use of nudging to alert a smart phone user about the possible privacy leaks~\cite{lorrie2014nudge}, a technique which can readily be applied to warn users of the dangers of taking selfies in the present location/situation.

In this work, we study the dangerous impacts of clicking a selfie. Our work is the first in trying to characterize all the selfie deaths that have occurred in the past couple of years. Till now, there has been no research that proposes features and methods to identify dangerous and non-dangerous selfies posted on social media, which is what we propose to do in this work.

\section{Selfie Deaths Characterization}
\label{sec:motivation}
In our work, we define a selfie-related casualty as \textit{a death of an individual or a group of people that could have been avoided had the individual(s) not been taking a selfie}. This may even involve the unfortunate death of other people who died while saving or being present with people who were clicking a selfie in a dangerous manner. 
To be able to better understand the reasons behind selfie deaths, victims, and such incidents, we collected every news article reporting selfie deaths. We used a keyword based extensive web searching mechanism to identify these articles~\cite{stringhini2013follow}. Further, we only considered those articles as credible sources which were hosted on the websites having either their Global Alexa ranking less than 5,000, or having a country specific Alexa rank less than 1,000. The earliest article reporting a selfie death that we were able to collect was published in March 2014. Two annotators manually annotated the articles to identify the country, the reason for death, the number of people who died, and the location where the selfie was being taken.

\begin{table}[!hbtp]
\centering
 \begin{tabular}{|>{\raggedright}p{3.5cm}|c|}
 \hline
\textbf{Country} & \bigcell{c}{\textbf{Number of Casualties} \\ (N=127)}	\\	\hline
India & 76	\\	\hline
Pakistan & 9	\\	\hline
USA & 8	\\	\hline
Russia & 6	\\	\hline
Philippines, China & 4	\\	\hline
Spain & 3	\\	\hline
Indonesia, Portugal,
Peru, Turkey & 2	\\	\hline
Romania, Australia, Mexico, 
South Africa, Italy, Serbia,	
Chile, Nepal, Hong Kong & 1		\\	\hline

 \hline
 \end{tabular}
 \caption{Country-wise number of selfie casualties}
\label{table:country_selfiedeaths}
\end{table}

Using our approach, we were able to find $127$ selfie-related deaths since March 2014. These deaths involved $24$ group incidents, and others were individual incidents. By group incidents, it is meant that multiple deaths were reported in a single incident. An example of this could be an incident near Mangrul lake in the Kuhi district in India, where a group of $10$ youth had gone for boating in the lake. While they were trying to take selfie, the boat tilted, and $7$ people died. We count all such incidents as group incidents. Out of all the group incidents, $16$ of the incidents involved $2$ individuals, $5$ involved $3$ people, $1$ incident had $5$ casualties, and there were $2$ group incidents claiming the lives of $7$ people each. By analyzing selfie deaths - in terms of group and individual deaths, it can be concluded that taking dangerous selfies not only puts the selfie-taker at a risk but also can also be hazardous to the people around them. Although it is known that women take more selfies than men~\cite{women_men_selfies}, however, our incident analysis showed that men are more prone to taking dangerous selfies, and accounted for roughly $75.5$\% of the casualties. Out of all the deaths, $41$ victims were aged less than $20$ years, $45$ were between $20$ and $24$ years of age and $17$ victims were $30$ years old or above. This is consistent with our earlier finding that the trend of taking selfies is really popular among millennials.

Studying the geographic trends of the selfie deaths, we observed that India accounted for more than $51.76 \%$ of the overall incidents, out of which $87\%$ were water-related casualties. In the USA, 3 deaths occurred while trying to click a selfie with a weapon, followed by Russia with 2 casualties. This might be a consequence of the open gun laws in both the countries. Distribution of incidents according to the country is shown in Table~\ref{table:country_selfiedeaths}.


\begin{figure}[!ht]
    \centering
    \includegraphics[scale=0.15]{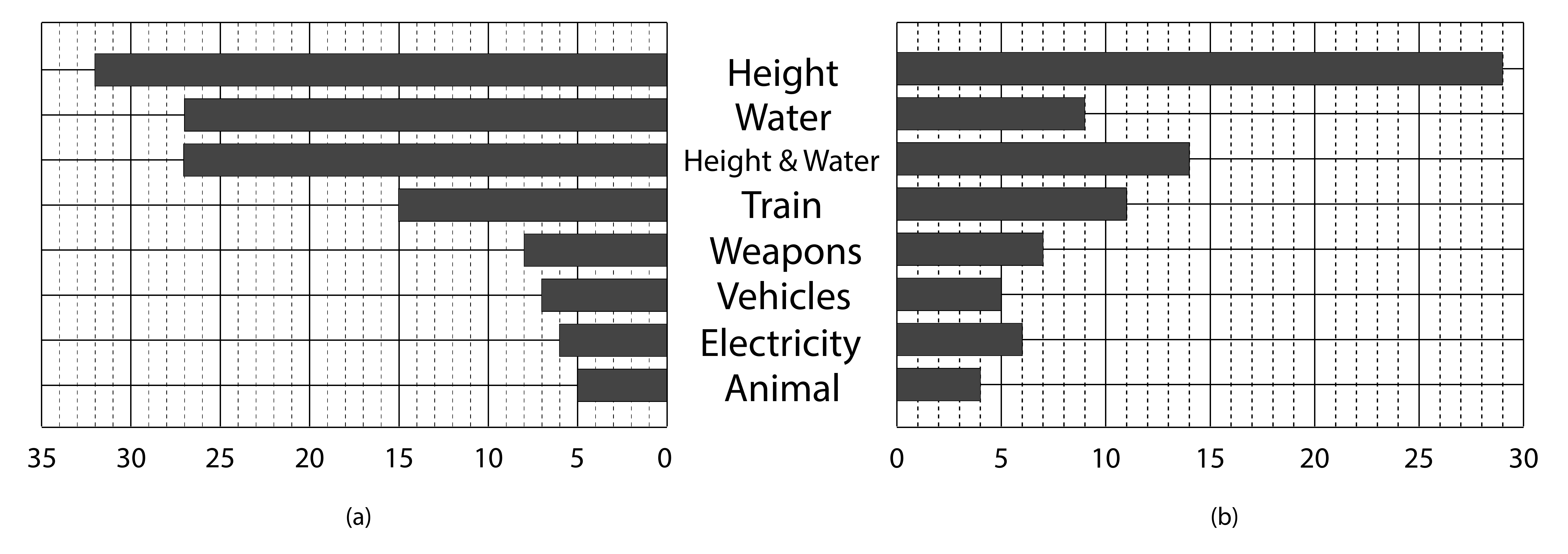}
    \caption{(a) Number of Deaths due to various reasons, and (b) Number of Incidents.}
    \label{fig:double_trouble}
\end{figure}

We looked at all the articles in our database to figure out what are the most common factors/reasons behind selfie deaths. Overall, we were able to find $8$ unique reasons behind the deaths. We found that most common reason of selfie death was height-related. These involve people falling off buildings or mountains while trying to take dangerous selfies. Figure~\ref{fig:double_trouble} shows the number of casualties for various reasons of selfie deaths. From the plot, it can be observed that for water-related causes, there were more group incidents. There were also considerable number of incidents where the selfie-taker exposed himself to both the height related and water body related dangers, thus we have analyzed such incidents separately. Twenty-seven individuals who died in $14$ incidents qualified for this category. The second most popular category was being hit by trains. We found that taking selfies on train tracks is a trend. This trend caters to the belief that posting on or next to train tracks with their best friend is regarded as romantic and a sign of never-ending friendship.\footnote{http://www.dw.com/en/dangerous-trend-the-train-track-selfie/a-18932440}

After analyzing selfie deaths, we can claim that a dangerous selfie is the one which can potentially trigger any of the above-mentioned reasons for selfie deaths. For instance, a selfie being taken on the peak of a mountain is dangerous as it exposes the selfie taker to the risk of falling down from a height. To be able to warn more users about the perils of taking dangerous selfies, it is essential to have a solution that can distinguish between the dangerous and non-dangerous selfies. Motivated by the reasons that we found for selfie deaths, we formulated features which would be ideal to provide enough differentiation between the $2$ categories. In future sections, we discuss in detail as to how we generated features for different selfie-related risks and develop the classifier to identify selfies that are potentially dangerous.


\section{Selfie Dataset Curation}
\label{sec:dataset}
We used \textit{Twitter} for our data collection. Twitter is a popular social media website which allows access to the data posted by its users through APIs. Twitter provides an interface via its \textit{Streaming API} to enable researchers and developers to collect data.\footnote{\url{https://dev.twitter.com/streaming/overview}} Streaming API is used to extract tweets in real-time based on the query parameters like words in a tweet, location from where the tweet is posted and other attributes. The API provides $1$\% sample of the entire dataset~\cite{morstatter2013sample}. We collected tweets related to selfies using keywords like \textit{\#selfie}, \textit{\#dangerousselfie}, \textit{\#extremeselfie}, \textit{\#letmetakeaselfie}, \textit{\#selfieoftheday}, and \textit{\#drivingselfie}.
We collected about $138$K unique tweets by $78$K unique users. The descriptive statistics of the data are given in Table~\ref{table:twitterdata_selfies}.

\begin{table}[!hbtp]
\centering
 \begin{tabular}{|l|c|}
 \hline
\textbf{Total Tweets} & 138,496 \\	\hline
\textbf{Total Users} & 78,236	\\	\hline
\textbf{Total Tweets with Images} & 91,059	\\	\hline
\textbf{Total Tweets with geo-location} & 9,444	\\	\hline
\textbf{Total Tweets with Text besides Hashtags} & 112,743	\\	\hline
\textbf{Time of first Tweet in our Dataset} & Mon Aug 01 	\\	\hline
\textbf{Time of last Tweet in our Dataset} & Tue Sep 27 \\	\hline
 \end{tabular}
 \caption{Descriptive statistics of Dataset collected for Selfies}
\label{table:twitterdata_selfies}
\end{table}

Out of the 138,496 tweets collected, we only found 91,059 to have images in them. We consider only those tweets for further analysis. However, it is not clear if all of those images were actually selfies or not. To retain only the true selfie images, we build a classifier based on image features to retain only the images that are selfies. We explain the classifier used below.

\textbf{Preprocessing:}
We manually annotated 2,161 images as to determine whether they were selfies or not. Out of the tagged images, we found that 1,307 (roughly $60$\%) were selfies, and remaining 854 were not selfies. Using the manual annotations as ground truth, we constructed a classifier to discriminate between the selfies and non-selfies.
The classifier was based on the transfer learning based model called DeCAF proposed by Donahue et al.~\cite{Donahue2014}. DeCAF model first trains a deep convolutional model in fully supervised setting, and then various features from this network are extracted and tested on generic vision tasks. The deep convolutional model is as mentioned in Szegedy et al.~\cite{SzegedyImageNet}. The convolutional model has been trained and tested on the task of classifying $1.2$ million images in ImageNet LSVRC - $2010$ contest into 1,000 classes. It obtained top-1 and top-5 error rates as $21.2$\% and $5.6$\% respectively. As specified in the DeCAF framework, we use this trained model for the task of identifying if an image is a selfie image or not. This approach is useful as the cost of annotating all images as to whether it is a selfie or not is saved, and most convolutional deep learning models require enormous amounts of training data to train effectively from scratch. Therefore by using DeCAF, we built on the generic features provided by the original convolutional neural network. We found that algorithm gave $88.48$\% accuracy with $10$-fold cross validation.

Using the model trained on the annotated dataset, we obtained labels for all of the non-annotated images. We found that out of 90K images (tweets with images or tweets hyper-linking to images), 62K were actually selfies. These 62K tweet set contained only 6,842 tweets which had a geolocation.

\section{Feature Set Generation}
\label{sec:exp}
In this section, we discuss the features we use for our classifier to differentiate between dangerous and non-dangerous selfies.
Based on the analysis of selfie casualties we did in Section~\ref{sec:motivation}, we design different features for every major possible selfie-related risk (see Figure~\ref{fig:double_trouble}). We analyze each of the possible causes and consider what all features are possible in terms of tractability and availability. We first review the location-based features.

\textbf{Height Related Risks:}  From our dataset, we observed that $29$ selfie deaths were because of falling from an elevated location. We take this as an indication that taking selfies at an elevated location is dangerous. Based on the location of the selfie, we want to generate features that tell us if an image has been taken at an elevated location or not. To estimate the elevation of a location, we used Google Elevation API.\footnote{\url{https://developers.google.com/maps/documentation/elevation/}}

Taking only the elevation of a particular place is not be informative to tell if the location is actually dangerous or not. For example, if a city is at a higher altitude, that does not make it necessarily dangerous. However, sudden changes in the nearby terrain indicate that there is a steep decrease in elevation, making the location dangerous. Google Elevation API returns negative values for certain locations such as water body. We formulated the following features based on the elevation of the location:


\begin{figure*}
\centering
\begin{tabular}{lcl}
\scalebox{0.20}{\includegraphics[angle=0,
origin=c]{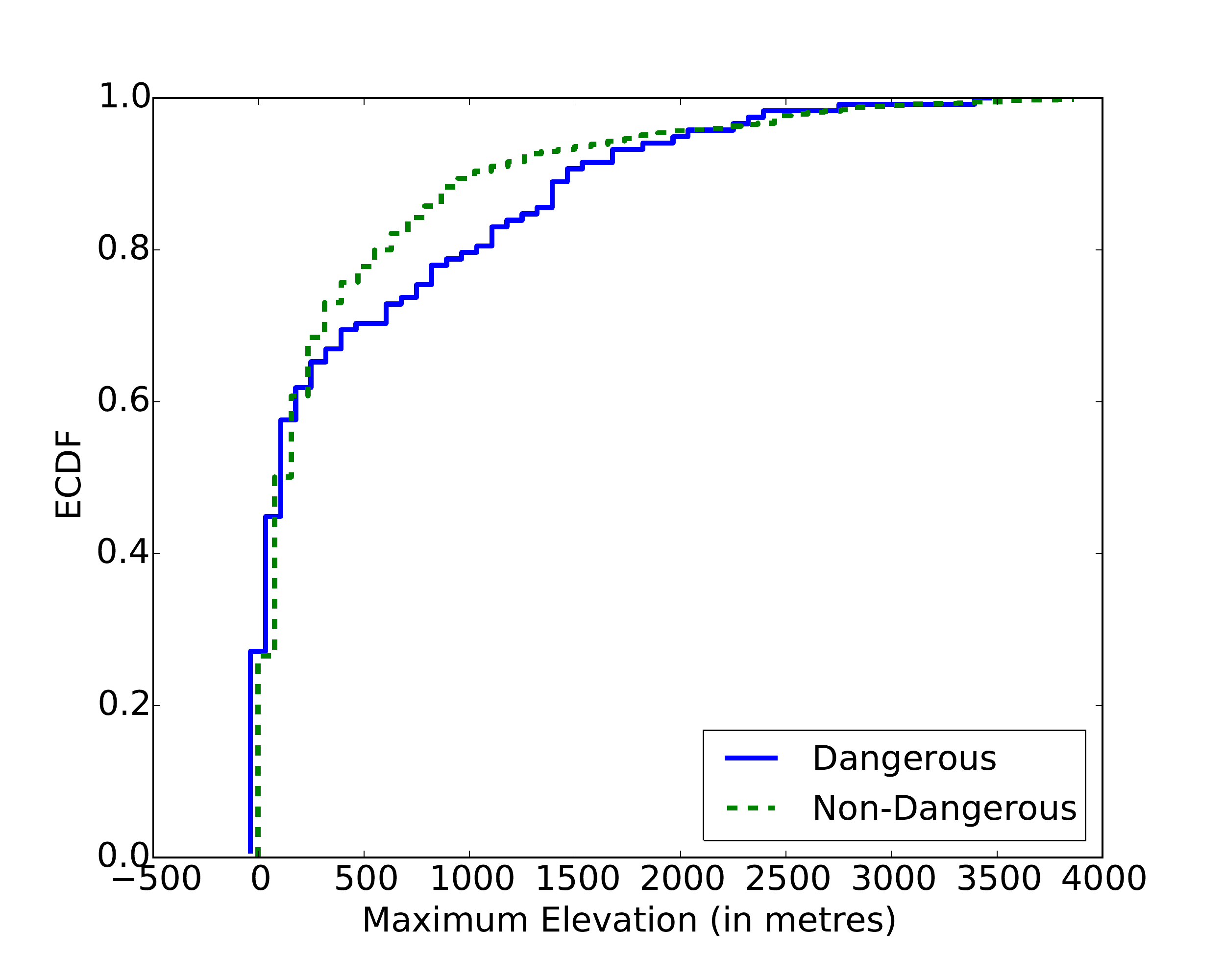}} &
\scalebox{0.20}{\includegraphics[angle=0,
origin=c]{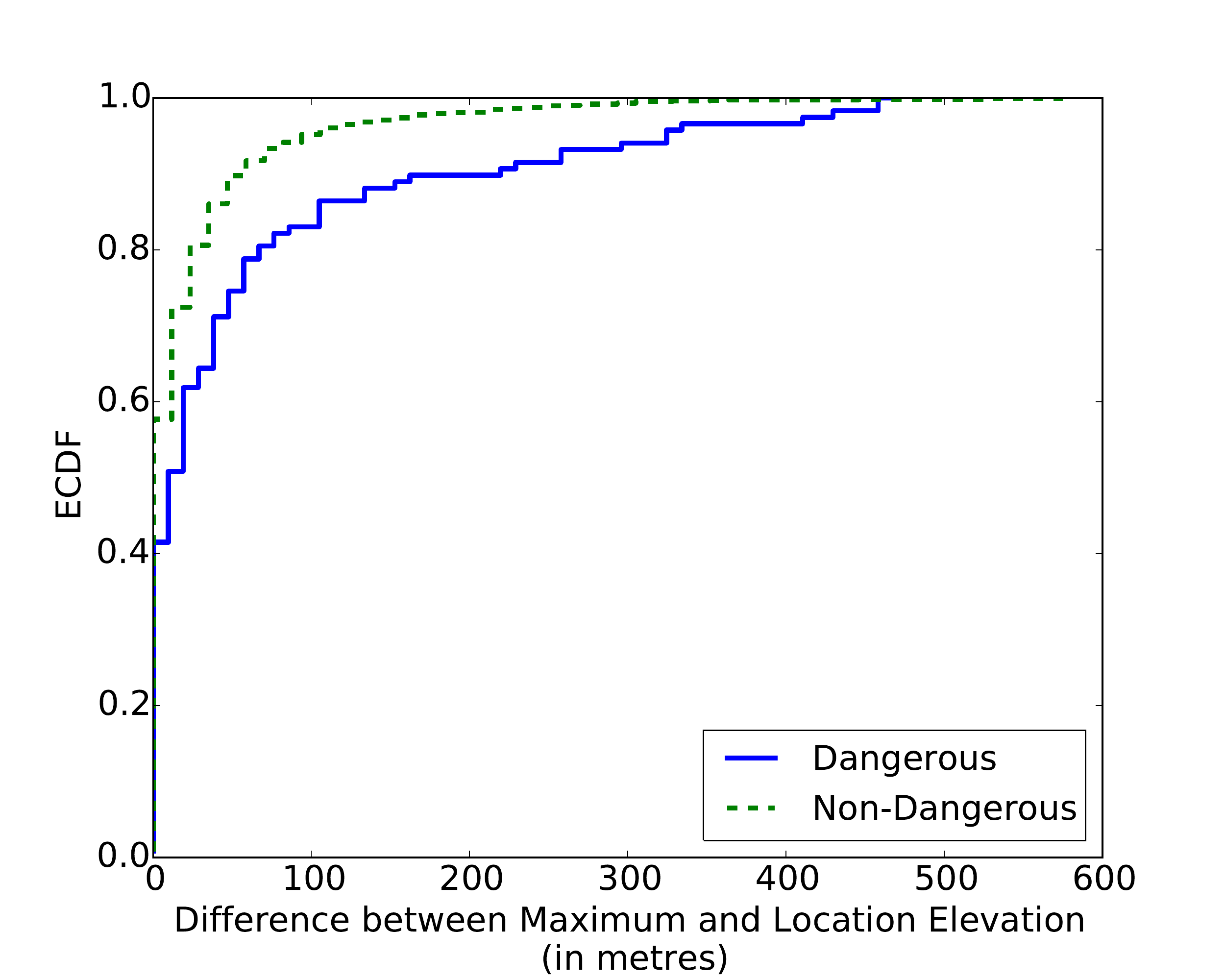}} &
\scalebox{0.20}{\includegraphics[angle=0,
origin=c]{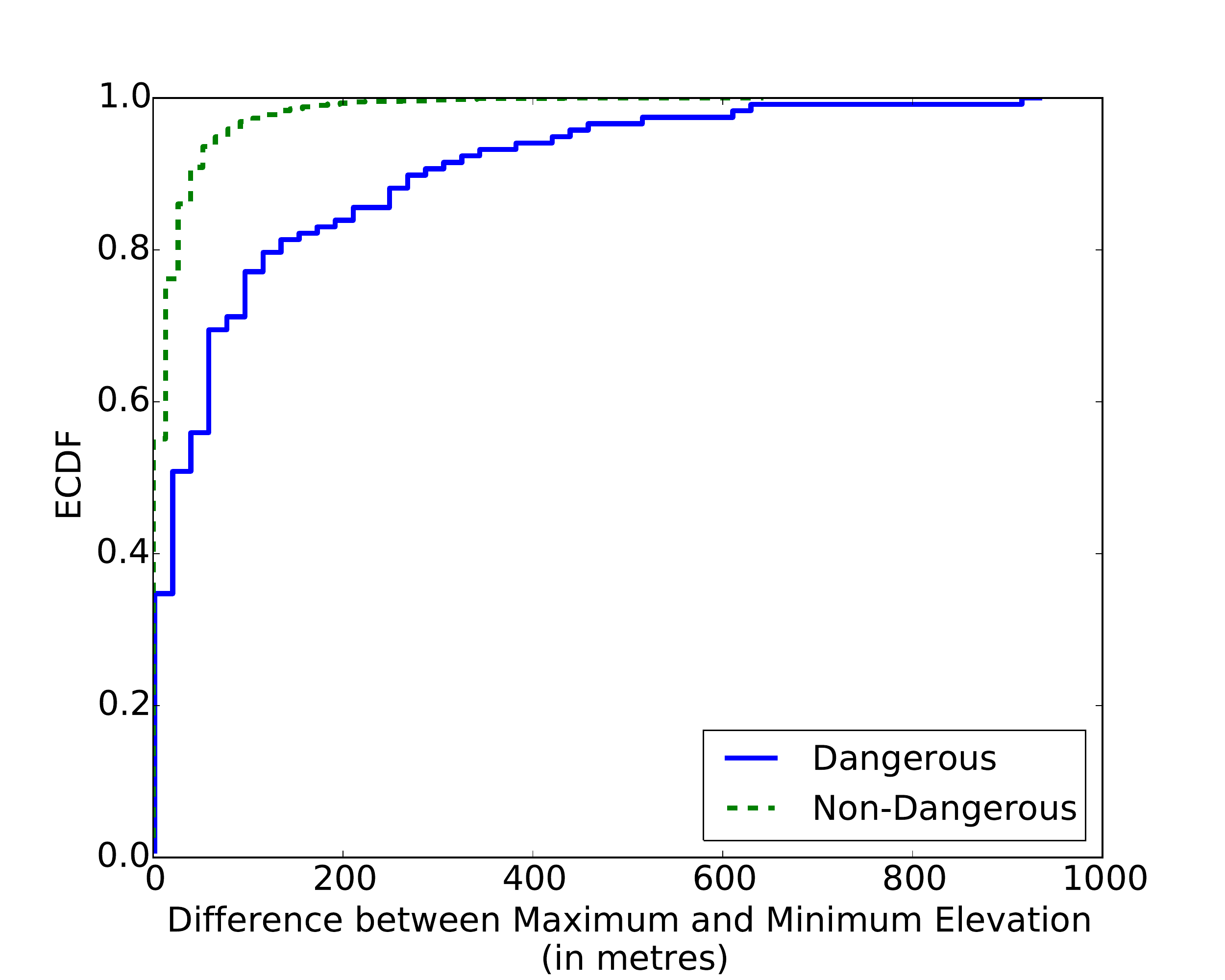}}
\end{tabular}
\caption{CDF Plots showing the difference in the distribution of height-related features for dangerous and non-dangerous images. Left: Maximum Elevation in 5km radius and 5 sampled locations (p-value:0.028). Center: Maximum difference in elevation of $10$ points sampled in $1$km radius with the elevation of the location (p-value: 7.09e-6). Right: Maximum Elevation Difference of $10$ points sampled in $1$km radius (p-value: 1.22e-9).}
\label{fig:cdf_height}
\end{figure*}

\begin{itemize}
\item{\textit{Elevation of the exact location of the selfie:} This feature was not informative as it captures only the elevation of the location, and that does not necessarily mean a risk due to height. This was validated by the fact that p-value of Kolmogorov-Smirnov (KS) $2$ sampled test was $0.12$; which we can reject only in $15$\% confidence interval.}
\item{\textit{Maximum Elevation of the surrounding area:} To get a sense of the area surrounding the exact location, we sample $10$ locations in $1$-km radius and return the maximum elevation out of those. We choose the specified value of radius and number of locations because they returned the lowest p-value after applying $2$-sample KS test for dangerous and non-dangerous selfie distribution.}
\item{\textit{Difference Elevation of the surrounding area:} We calculate this as the maximum difference between the elevation of our exact location and the sampled locations' elevation. These features capture the sudden elevation drop that might exist near the surrounding area. For this feature, we sampled $5$ locations in a $5$-km radius for the same reason as mentioned above.}
\item{\textit{Maximum Elevation Difference in the surrounding area:} Taking the maximum difference between the highest elevation and lowest elevation of the sampled points helped us capture the amount of elevation variation in the surrounding area.}
\end{itemize}
We did not work with other possible statistics such as the average elevation or median elevation as those statistics try to capture the center point or a single representative value of the distribution. We are however interested in sudden elevation drops in the surrounding area, which will lie on the extremes of the elevation distribution.

To evaluate the efficiency (or the discriminative power) of the above-mentioned features, we plot the empirical cumulative distributions (CDF) of height-related dangerous selfies and non-dangerous selfies. This can be seen in Figure~\ref{fig:cdf_height}. We can notice that for the $3$ features, the empirical CDF of dangerous and non-dangerous selfies are considerably different. The KS test returned p-values:0.028 for Maximum elevation, 7.09e-6 for Elevation difference between maximum elevation and our location and 1.22e-9 for Maximum elevation difference.



\begin{figure}[!ht]
\centering
\subfigure[]{
\includegraphics[scale=0.15]{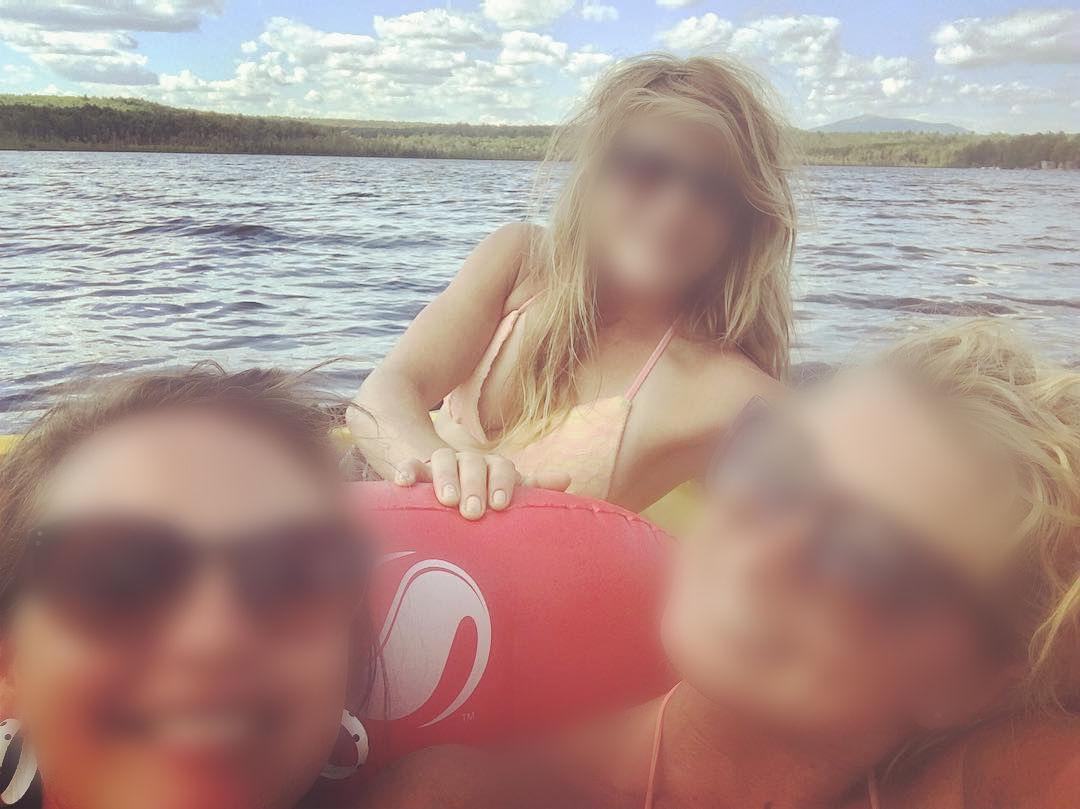}
}
\subfigure[]{
\includegraphics[scale=0.25]{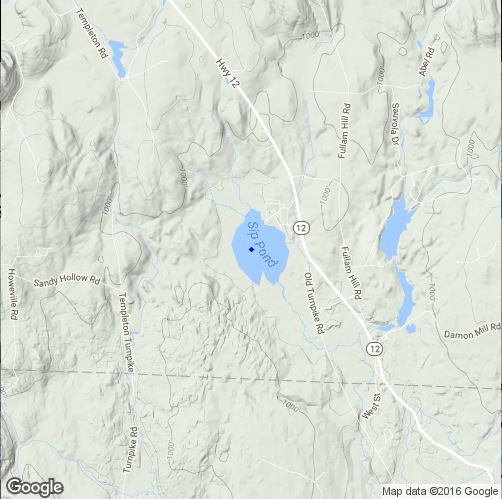}
}
\subfigure[]{
\includegraphics[scale=0.25]{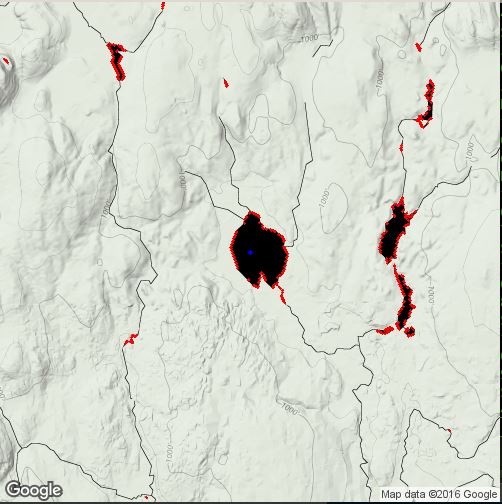}
}
\caption{Segmentation Example: Different Stages of processing to get the final segmented image distinguishing between the water and land.}
\label{fig:waterseg}
\end{figure}

\textbf{Water Related Risks:} Another prominent reason of selfie casualties that we infer from Figure~\ref{fig:double_trouble} is water-related risks. After analyzing the water-related incidents, we found that often people took selfies while being in a water body or in close proximity to one. They ended up drowning by losing their body balance and falling into the water body. To tackle water related risks, we generate features based on the proximity of their location to a water body. Consider the selfie in Figure~\ref{fig:waterseg}(a) which has been taken in the middle of a water body. We mapped the exact location of the selfie to Google Maps and considered $500 \times 500$ pixel image pertaining to level $13$ zoom factor on Google Maps \cite{google_zoom}. The image after this step looked like in Figure~\ref{fig:waterseg}(b). We applied image segmentation to identify the contour of all the water bodies shown in Figure~\ref{fig:waterseg}(c). To infer whether a given location is in close proximity to a water body or not, we use the minimum distance to a water body from the location of the image as a feature. Since all the segmented images were of maps with same scale and zoom factor, the distance was treated as pixel location distance. Proximity to a small water body like a stream or a river might not make a selfie dangerous, therefore we also use fraction of the pixels in the segmented image (Figure~\ref{fig:waterseg}(c)) to further help us in distinguishing between dangerous and non-dangerous selfies.

\begin{figure}[!ht]
\centering
\begin{tabular}{ll}
\scalebox{0.16}{\includegraphics[angle=0,
origin=c]{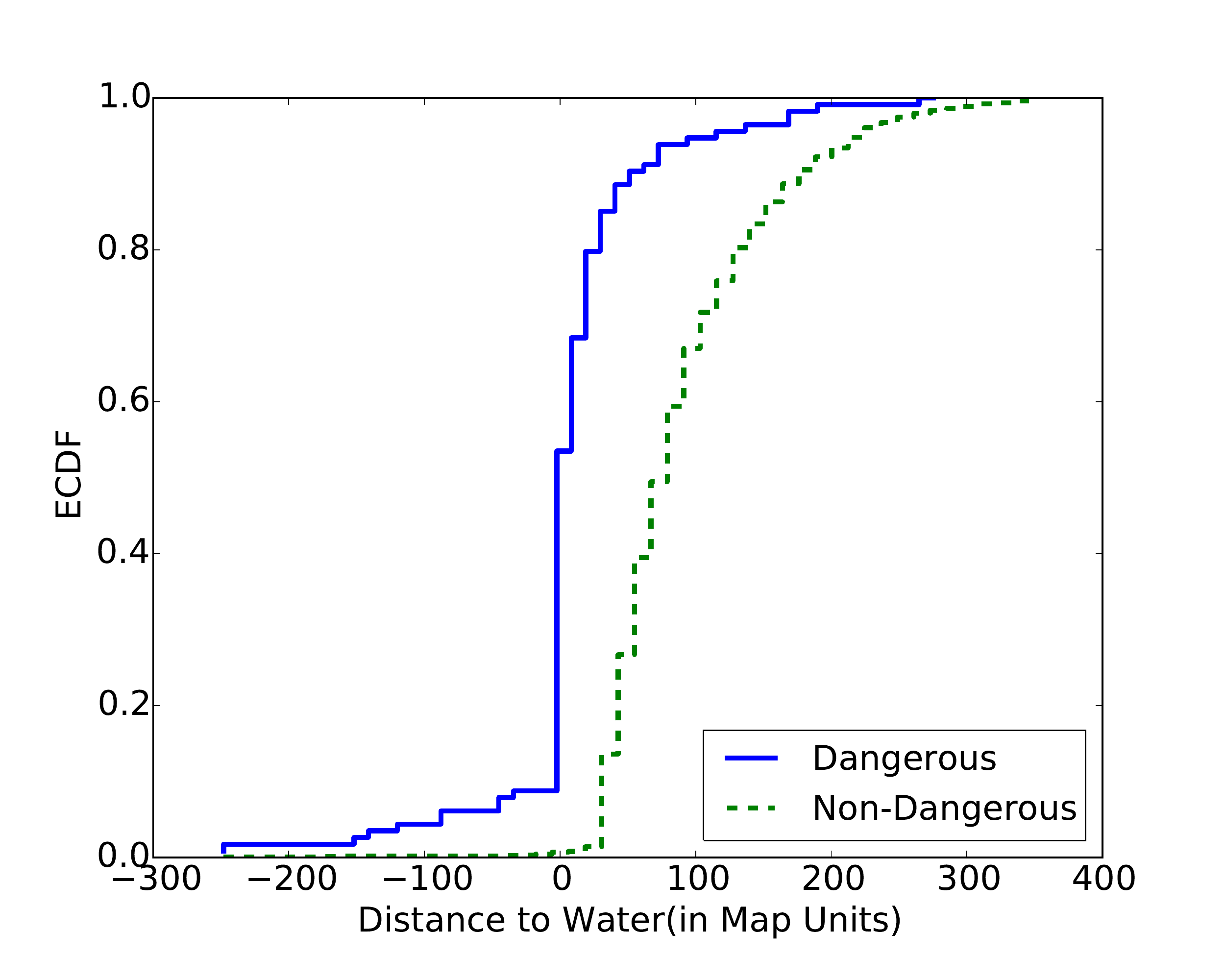}} &
\scalebox{0.16}{\includegraphics[angle=0,
origin=c]{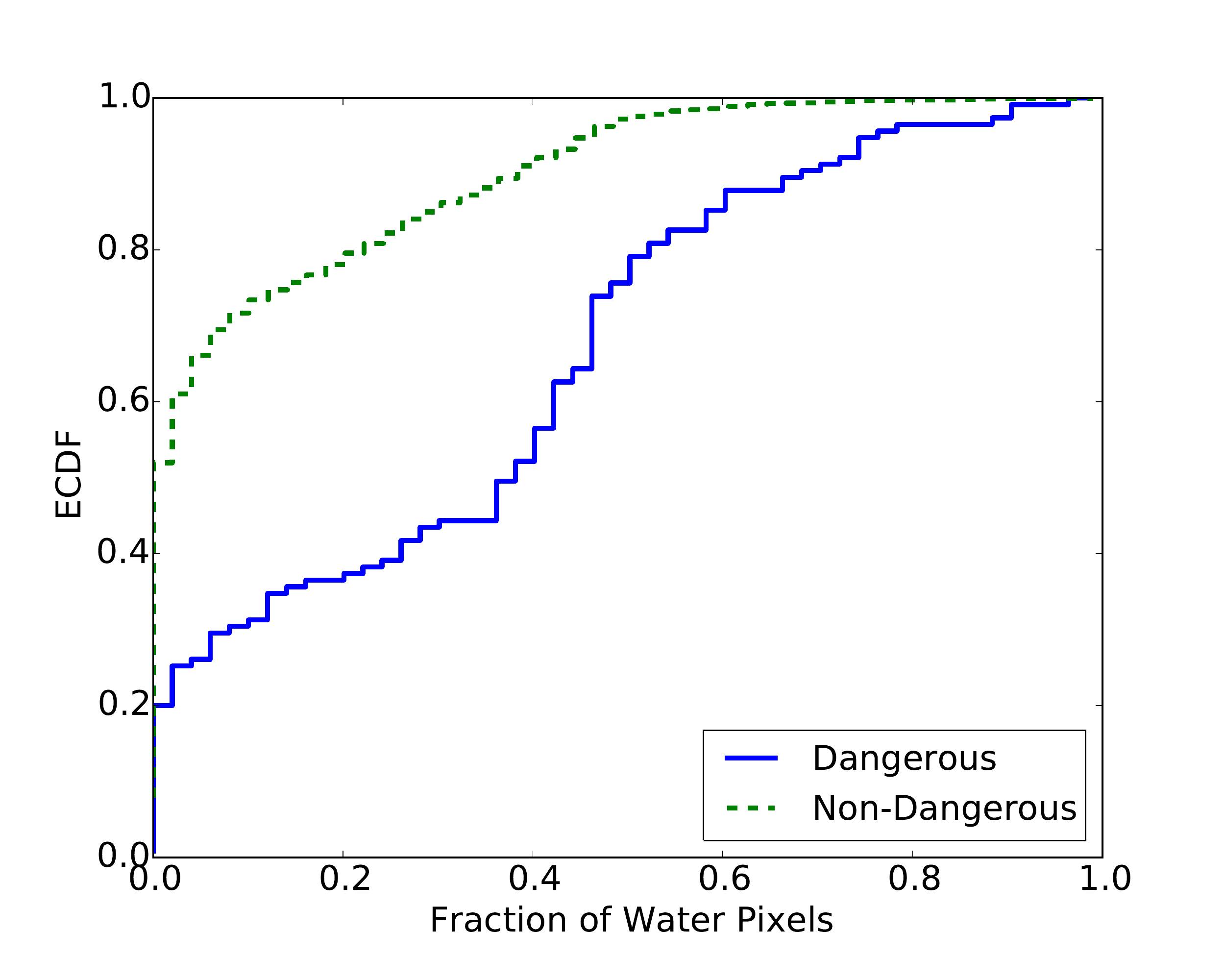}}
\end{tabular}
\caption{CDF Plots showing the difference in dangerous and non-dangerous distributions for water-related features. Left: Minimum distance to a water body. Right: Fraction of water pixels in the segmented image}
\label{fig:water_cdf}
\end{figure}

We can observe from the Figure~\ref{fig:water_cdf} that for both of the water features - minimum distance to a water body and the fraction of water pixels in the segmented image, the distribution of water-related dangerous and non-dangerous selfies is considerably different. We use $2$-sampled KS test to statistically confirm our observations. We obtained p-values of 1.18$e$-19 (minimum distance to a water body) and 2.79$e$-19 (fraction of water pixels in the segmented image) indicating that we can safely reject that the features are being generated from the same distribution.

\textbf{Train/ Railway Related Risks:} Besides water and height-related risks, another common reason of selfie casualties is train-related risks which accounted for $11$ casualties. We used Google Places API to determine if there is a railway track or a railway station close to the location of the selfie or not. We used the minimum distance between the location and the railway track as a feature. Though this feature is not sufficient to distinguish between dangerous and non-dangerous selfie, it still provides valuable information which when appended to other features proves to be helpful in the classification task.

\textbf{Driving/Road Related Risks:} It is challenging to account for driving-related risks in all possible contexts. The location of the selfie can provide information about how close a person is to a road. Using only the location data is not sufficient to determine if the selfie-taker was driving at the time of taking a selfie, or was standing in the middle of a busy road to take the selfie. However, we still think that the minimum distance of the location of the selfie to the highway/road will be informative in determining the `dangerousness' of the selfie when used in conjuction with other features.

For all the other reasons such as weapons, animal, electricity, it is difficult to find location based insights, and thus impossible to find location based features. We rely on other signals based on the text accompanying the selfie, and the content of the image to be able to derive features which can provide insights about these reasons. For example, the presence of a weapon or animal can be easily inferred from the image content. Below, we discuss the text-based and image content-based features.

\textbf{Text-based Features:}
The content of the tweet can be a useful source for indicating if the image accompanying it is a dangerous selfie. Users tend to provide context to the image either directly in the tweet text or through hashtags. We use both to generate our text-based features. After removing the URLs, tokenizing the tweet content, and processing emojis, we obtain our text input. We use TF-IDF over the set of unigrams and bigrams. For further enriching the text feature space, we convert the text into a lower dimension embedded vector obtained using \textit{doc2vec}\cite{doc2vec}.

\textbf{Image-based Features:}
Since an image could be dangerous due to various reasons, we cannot simply apply a classifier to the actual pixels of the image. Classifying an image as to whether it is dangerous or not requires more understanding of the context and the elements in the image. Therefore, we first extract the salient regions in images and then generate captions for each of those regions.

\begin{figure}[!ht]
\centering
\begin{tabular}{ll}
\scalebox{0.16}{\includegraphics[angle=0,
origin=c]{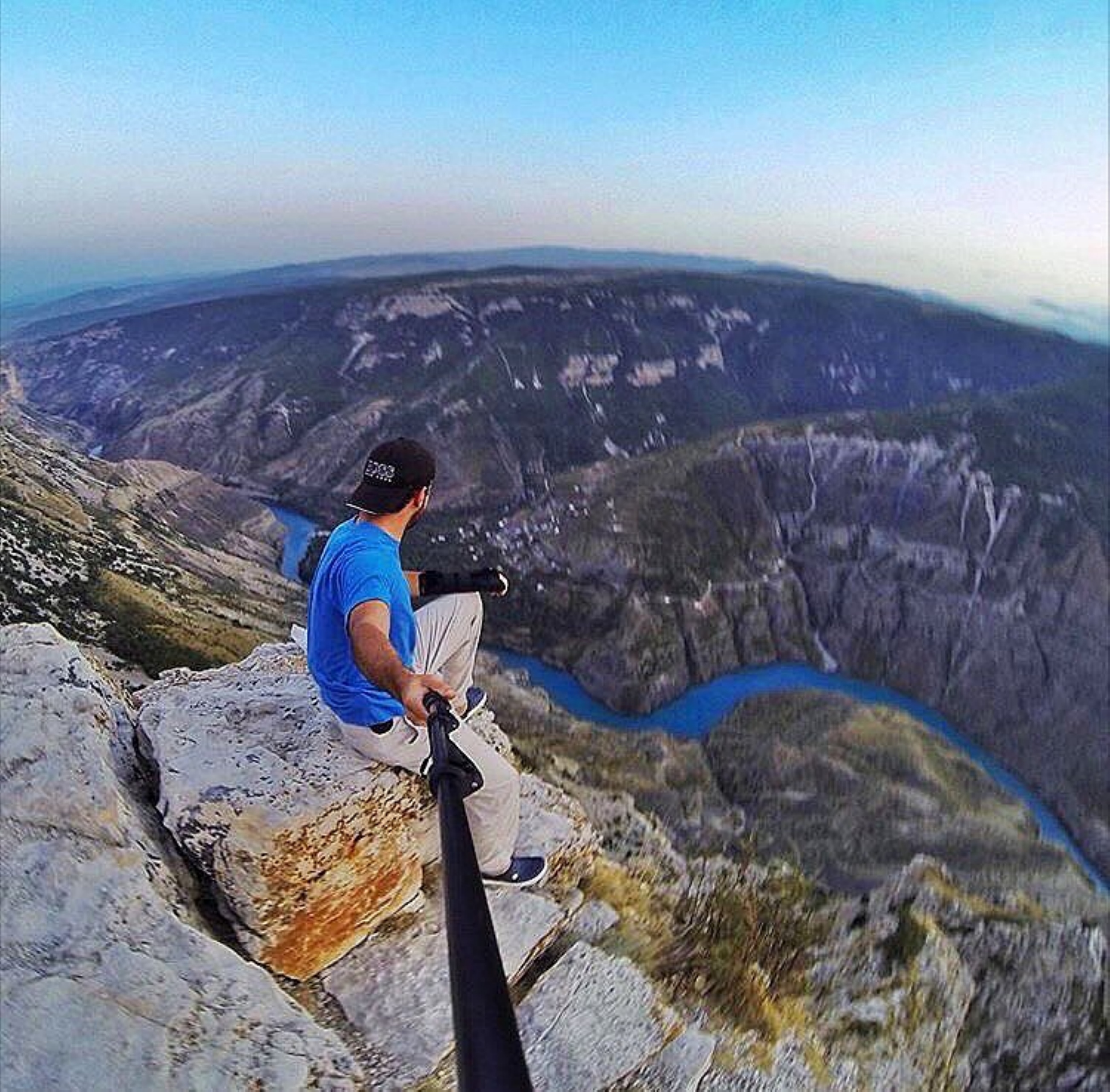}} &
\scalebox{0.25}{\includegraphics[angle=0,
origin=c]{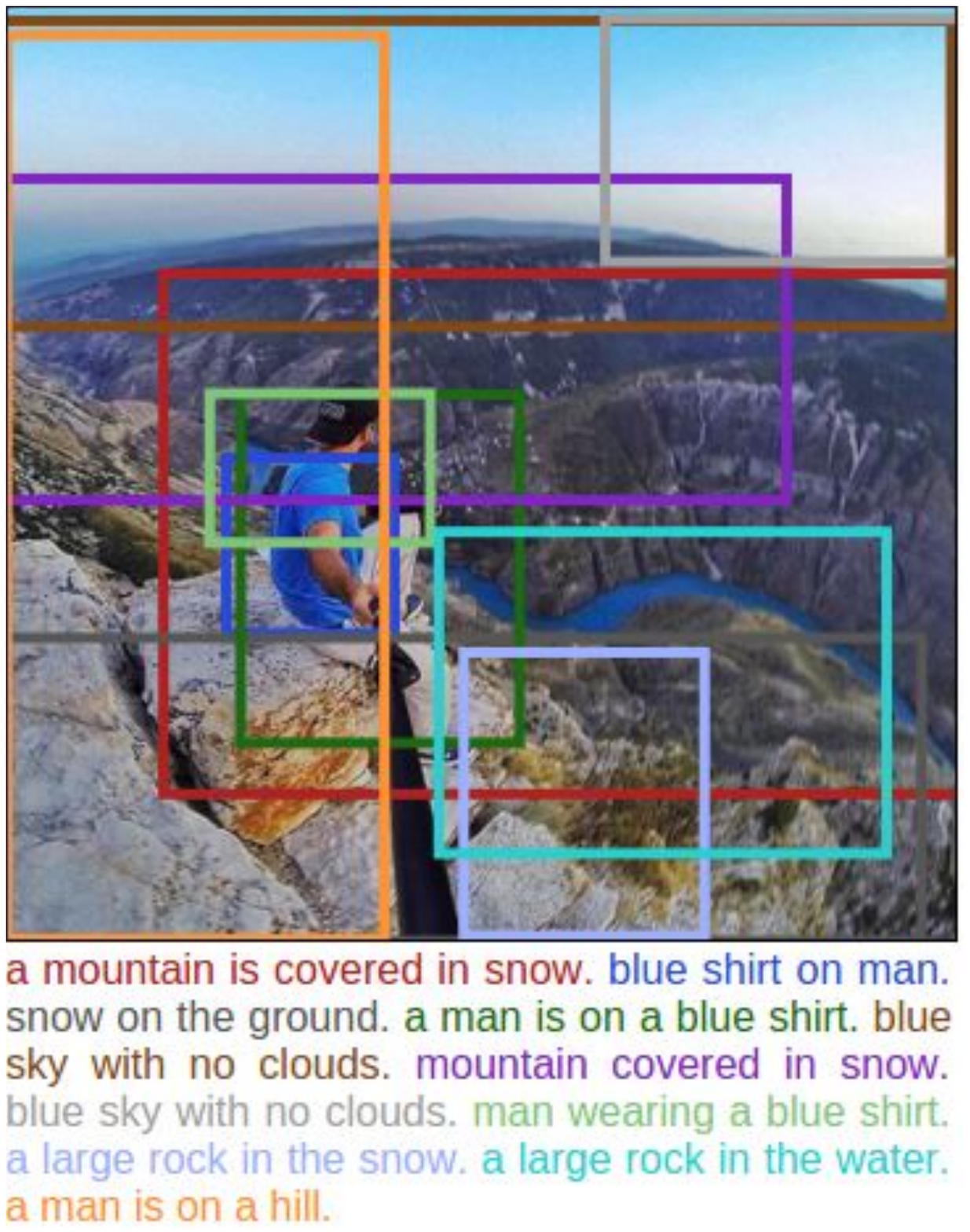}}
\end{tabular}
\caption{An example of the DenseCap on one of the images (Left) from our dataset. We use the dense captions produced by DenseCap (Right) to come up with text based features over them.}
\label{fig:densecap_example}
\end{figure}

To extract informative regions in images and for the caption-generating process, we used DenseCap~\cite{densecap}. DenseCap is start-of-the-art deep learning based captioning technique for regions in an image. It outperforms other models such as Full Image RNN, Region RNN on both tasks of dense captioning and as well as image retrieval comfortably. The average precision on the dense captioning task by DenseCap was $5.24$, way higher than the closest competitor $4.88$. The architecture of DenseCap involves a fully convolutional layer, a fully convolutional localization layer used for extracting ROI (regions of interest) and their features, a recognition network for finding relevant ROI's, and a language model to generate captions for the ROI. An example of the output of the DenseCap on a selfie in our dataset is shown in Figure~\ref{fig:densecap_example}.

\begin{figure}[!ht]
\centering
\includegraphics[scale=0.3]{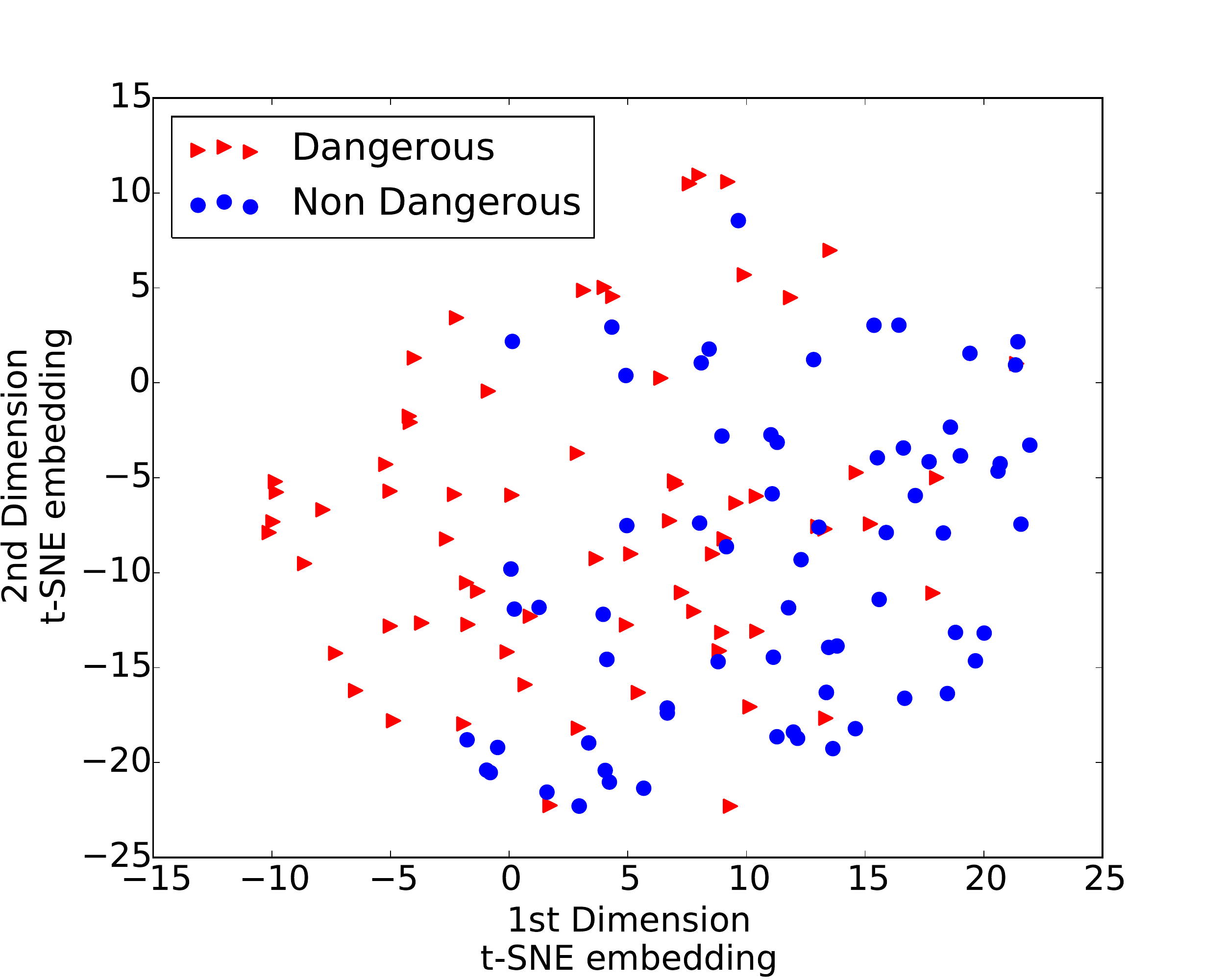}
\caption{t-SNE scatter plot of doc2vec output of generated captions for $50$ randomly chosen dangerous and non-dangerous selfies.}
\label{fig:densecap_tsne}
\end{figure}

We treat the generated captions as the text describing the image in natural language. From the text, we compute natural language features such as unigrams, bigrams to determine if the content of the image is dangerous or not. We also convert the captions generated into a lower dimension vector in a similar fashion we did for text-based features. To empirically view the validity of our approach, we plotted the $2$-dimensional t-SNE (Stochastic Neighbor Embedding)~\cite{tsne} mapping of the embedded \textit{doc2vec} vectors in Figure~\ref{fig:densecap_tsne}. In the plot, we can see that the triangles (dangerous selfies) are negative in the 1st vector components (X-axis), whereas the circles (non-dangerous selfies) are largely positive. On the plot, we can imagine a line easily separating most of the dangerous and non-dangerous selfies.
Our entire feature space could be categorized as shown in Table~\ref{table:feature_set}.

\begin{table}[!ht]
\centering
 \begin{tabular}{|l|p{4.75cm}|}
 \hline
\textbf{Feature Type} & \textbf{Feature} 	\\	\hline
\multirow{6}{*}{Location Based Features} & Elevation of the location\\
& Maximum Elevation \\	
& Difference between Maximum elevation out of sampled points and elevation of the location.\\
& Maximum elevation difference in the set of sampled points	\\
& Minimum Distance to water body\\
& Fraction of water pixels in the segmented image 	\\
& Distance to railway tracks	\\
& Distance to major roadway/highway	\\
\hline
\multirow{3}{*}{Image Based Features} & TF-IDF of unigrams and bigrams on DenseCap captions\\
& Doc2Vec representation of DenseCap captions\\
\hline
\multirow{2}{*}{Text Based Features} & TF-IDF of unigrams and bigrams on the Twitter text\\
& Doc2Vec representation of Twitter text\\
\hline
 \end{tabular}
 \caption{Location-based, Image-based and Text-based features used for classification of selfies.}
\label{table:feature_set}
\end{table}


\section{Experiment}
\label{sec:exp}
\subsection{Manual Annotation}
\begin{figure}[hbtp!]
    \centering
    \includegraphics[scale=0.20]{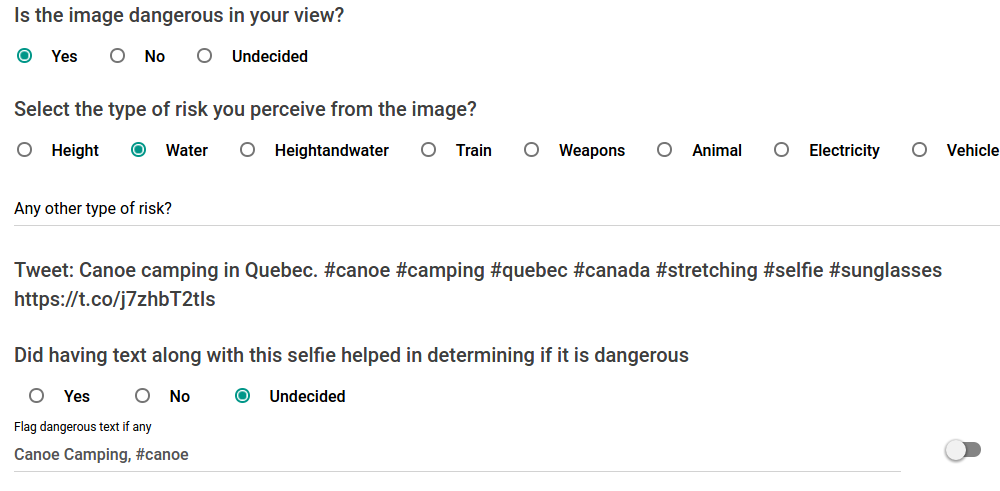}
    \caption{Screenshot of the annotation tool. We asked above questions to the annotators based on a selfie image shown to them.}
    \label{fig:annotation_tool}
\end{figure}

From the selfie data set described in Section~\ref{sec:dataset}, we sampled a random set of 3,155 selfies with geolocation for creating an annotated data set. We manually labeled the images to determine whether they are dangerous or not. For the process of annotations, we asked questions such as, whether the image depicted is dangerous or not? If yes, then what is the possible reason for it being dangerous? And, whether text accompanying the image helped them in classifying if image is dangerous or not, and so on. A screenshot of the tool is shown in Figure~\ref{fig:annotation_tool}.\footnote{The annotation tool we used is available at \url{http://twitdigest.iiitd.edu.in:4000}}
We asked $8$ annotators to annotate the set of 3,155 selfies, randomly split into a common set having $400$ images. The common set was annotated by every annotator, and the shared set was divided equally among all the annotators. The inter-annotator agreement rate obtained on the common set of $400$ selfies, using the Fleiss Kappa metric~\cite{fleiss1971measuring} was $0.74$. Fleiss kappa metric interpretation reveals that the above value indicates substantial agreement between the annotators~\cite{kappaintepretation}. The annotated dataset contained $396$ dangerous and 2,676 non-dangerous selfies. Annotators were unsure about the remaining selfies in our dataset.
For the annotated images, we found that vehicle related causes for a selfie being dangerous, like taking a selfie in a car, is the maximum, followed by water related risks. Statistics about the risks that annotators perceived from the dangerous images is given in Table~\ref{table:annotator_dangers}. Annotators frequently found images to be dangerous in more than one aspect. For such cases, we counted their labels for all the mentioned risk types. One striking observation is that even though we didn't find any selfie casualties due to road related incidents in our research, it was identified as a potential risk by the annotators in as many as 29 dangerous images~(7\%).

\begin{table}[!hbtp]
\centering
 \begin{tabular}{|l|c|}
 \hline
\textbf{Reason} & \textbf{Number of Dangerous Selfies}\\	\hline
Vehicle Related & 120	\\	\hline
Water Related & 118	\\	\hline
Height Related & 86	\\	\hline
Height and Water Related & 55	\\	\hline
Road Related & 29	\\	\hline
Animal Related & 16	\\	\hline
Train Related & 8	\\	\hline
Weapons Related & 4	\\	\hline
 \end{tabular}
 \caption{Reasons marked by annotators for a selfie being dangerous.}
\label{table:annotator_dangers}
\end{table}

\begin{table*}[!ht]
\centering
 \begin{tabular}{|l|c|c|c|}
 \hline
  & \textbf{Water Related Danger} & \textbf{Height Related Danger} & \textbf{Vehicle/Road Related Danger}	\\	\hline
  \textbf{Accuracy} & \textbf{0.851} & 0.773 & 0.705	\\	\hline
  \textbf{Precision} & \textbf{0.873} & 0.81 & 0.738 	\\	\hline
  \textbf{Recall} & \textbf{0.851} & 0.801 & 0.714	\\	\hline
  \textbf{F1-Score} & \textbf{0.857} & 0.801 & 0.721	\\	\hline
  \textbf{Technique} & Random Forest & Random Forest & SVM \\	\hline

 \end{tabular}
 \caption{Performance of individual risk classifiers with 10-fold cross validation, along with the technique which yielded these results}
\label{table:expB_accuracy}
\end{table*}

\subsection{Classifier}
Considering the annotations performed in the section above as ground truth, we evaluate the performance of our classifier on the task of classifying whether a selfie is dangerous or not.
The problem of classifying dangerous selfies is a highly unbalanced problem. We have only 396 (roughly 13\%) dangerous selfies in comparison to the remaining 2.6K non-dangerous selfies. Therefore, we use random under-sampling to reduce the majority class samples (non-dangerous) such that the number of non-dangerous selfies is equal to the number of dangerous selfies. 

We divide the process of experimentation into two broad parts:

\subsubsection{Identifying Dangerous Selfies}
Using the features generated, we try to predict if a given selfie is dangerous or not. As shown in Table~\ref{table:feature_set}, our feature space can be easily classified into $3$ categories - text-based, image-based and location-based. To compare all of the feature types, we build and test the classifiers for every possible combination of the features. For all our experiments, we perform $10$-fold cross validation. Furthermore, we use grid search to find ideal set of hyperparameters for each classifier by doing a $3$-fold cross validation on the training set. We tested the performance of our method using $4$ different classification algorithms - Random Forests, Nearest Neighbors, SVM and Decision Trees. Each of the classifier was trained and tested on similar dataset and using the same feature configuration. Table~\ref{table:accuracy_classify} lists the accuracy obtained by using various classification techniques over different combinations of our feature space.

\begin{table*}[ht]
\centering
\begin{tabular}{|l|c|c|c|c|}
\hline & SVM & RandomForest & Nearest Neighbors & Decision Tree \\ \hline
\multicolumn{1}{|l|}{Image Only} & 0.72 & \textbf{0.73} & 0.55 & 0.67 \\ \hline
\multicolumn{1}{|l|}{Text Only} & \textbf{0.61} & 0.51  & 0.51 & 0.53 \\ \hline
\multicolumn{1}{|l|}{Location Only} & 0.58 & 0.56  & 0.56 & 0.57 \\ \hline
\multicolumn{1}{|l|}{Image + Location} & 0.70 & \textbf{0.72} & 0.55 & 0.64 \\ \hline
\multicolumn{1}{|l|}{Text + Location}  & \textbf{0.61} & 0.57  & 0.52 & 0.56 \\ \hline
\multicolumn{1}{|l|}{Text + Image} & \textbf{0.70} & \textbf{0.70} & 0.52  &  0.65 \\ \hline
\multicolumn{1}{|l|}{Text + Image + Location} & 0.68 & \textbf{0.73} & 0.54 & 0.65 \\ \hline
\end{tabular}
\caption{Average accuracy for 10-fold cross validation over different classification techniques and different feature configurations.}
\label{table:accuracy_classify}
\end{table*}

\textbf{\textit{Insight 1}}: We observe that image-based features consistently perform better than either of the text-based and location-based features. This is because the image-based features can capture the risk type which cannot be captured by location-based features, for example weapon-related or animal-related risks. Another reason is that the image-based features try to contextualize and infer meaning directly out of the image, and in a certain sense this is equivalent to our human annotators who have marked selfies as dangerous by looking at them and inferring whether they are dangerous or not visually.

\textbf{\textit{Insight 2}}: We applied 4 distinct machine learning classifiers - Random Forests, SVM (Support Vector Machines), Decision Trees and Nearest Neighbors. We noticed that Random Forests and SVMs performed consistently the best for all the given feature configurations. Random Forest, being an ensemble classifier has the property of reducing variance while not increasing the bias. It does so by training many individual decision trees on partitioned feature subspace. This also makes Random Forest robust towards high dimensional feature space. This is ideal in case due to the high dimensionality of the feature space.

\textbf{\textit{Insight 3}}: It can be noticed that all 3 features when combined give the highest accuracy. However, certain users might decide to not share location or might not have any text for their selfie, which might make it more challenging for the machine learning algorithms to classify if a selfie is dangerous or not. We observe that the features perform decently even in the absence of other features. The best feature type - image-based features perform with an accuracy of $73.6$\% set.

\subsubsection{Risk-Based Individual Classifier}
Besides trying to classify if a selfie is dangerous or not, we also wanted to test how well can we predict that a particular selfie is dangerous due to a particular reason. We used a similar methodology as mentioned in the above section for all our experiments. Out of the $8$ risks that were marked by the annotators and also inferred by characterizing selfie casualties, we developed classifier for $3$ categories - water related, height related and vehicle/road related. For the remaining categories, number of positive samples was insufficient to be able to train a classifier and more importantly, the generalizability of a classifier trained over such low number of samples will be doubtful. For a particular task, we used only those features which intuitively made sense to be used for predicting the given risk-type related dangerous selfies. An example of this could be that while predicting water-related dangerous selfie, it does not make sense to use height-based features or vehicle related features. The features space used for each risk type consists of image-based, text-based and location-based features. Location-based features consisted of features relevant to the risk type. To identify road-related dangerous selfies, we used the same location-based features that we used for vehicle/driving-related risks. 

We present the results for this experiment in Table~\ref{table:expB_accuracy}. We present the results for only the best configuration, and the best classifier. The best feature set for all $3$ tasks - water, height and vehicle related dangers was the combined space of all $3$ feature types - image-based, text-based and location-based. However, as mentioned earlier, the location-based classifier for every task was different and has been explained above.

\textbf{\textit{Insight 4}}: We were able to get better accuracy, precision statistics than the overall classifier for all the three tasks (water, height and vehicle) than the overall classifier discuss in the previous subsection. This is largely because we reduce the noise being added by other dangerous selfies that were dangerous because of different reasons, and had different distributions. Since most of the feature space was tuned to find a specific class of danger, it was hard for those features to be able to classify dangerous selfies for other distributions.

\textbf{\textit{Insight 5}}: The highest accuracy statistic was obtained on the water-task. This could be attributed to the design of features - minimum distance to water body and fraction of water pixels were easy to compute and unambiguous to indicate if a person is near to or in a waterbody. On manual investigation, we also found that DenseCap (source for image-based features) was able to identify water bodies in the selfies accurately. Moreover, the unambiguity in labeling water risks also helped.



%
%
%
%

%

%

%

\section{Discussion}			
\label{sec:concl}
%
%
In this paper we put forth a novel characterization of the selfie casualties that have occurred in the past. The rising trend of selfies and the dangers associated with careless selfie taking behaviour have been addressed in this paper. Our work helps in both understanding the various reasons behind selfie casualties and provides a potential solution to reduce such deaths. We presented a way to classify if a selfie image posted on the social media is dangerous or not. We used various classes of features such as - text-based features, image-based features and location-based features to represent different risk types. Location-based features were customized to capture the common reasons such as water-related, height-related reasons pertaining to selfie deaths. We used state of the art deep learning techniques such as DenseCap to get information about the content of the image to determine the nature of the selfie. We also tested the approach in the case of absence of one or more of the above mentioned features. We were able to identify dangerous selfies with an accuracy of $73$\%. Further, we also investigated if our feature space can form a classifier to predict a specific reason for the selfie being dangerous. We showed that we were able to identify water-related and height-related dangerous selfies with satisfactory accuracies.

Our classifier results are based on the human annotations and features that we learned from the selfie casualties. There is scope for improvement in the accuracy by increasing the dataset and the annotated dataset. The proposed methodology can help users know dangerous situations before taking a selfie. We hope to use our understanding from this paper to build a technology which can help users identify if a particular location is dangerous for taking selfies, and also provide information about casualties that have happened there in the past. We believe that the study can inspire and provide footprints for technologies which can stop users from clicking dangerous selfies, and thus preventing more of such casualties.

\bibliographystyle{abbrv}
\bibliography{BIB/other}


\end{document}